\documentclass[sigconf]{acmart}

\AtBeginDocument{%
  }

\copyrightyear{2026}
\acmYear{2026}
\setcopyright{cc}
\setcctype{by}
\acmConference[FORGE '26]{2026 IEEE/ACM Third International Conference on AI Foundation Models and Software Engineering}{April 12--13, 2026}{Rio de Janeiro, Brazil}
\acmBooktitle{2026 IEEE/ACM Third International Conference on AI Foundation Models and Software Engineering (FORGE '26), April 12--13, 2026, Rio de Janeiro, Brazil}
\acmPrice{}
\acmDOI{10.1145/3793655.3793716}
\acmISBN{979-8-4007-2477-0/2026/04}

\usepackage{tikz}
\usepackage{caption}
\usepackage{blindtext}
\usepackage{tcolorbox}
\usepackage[final]{pdfpages}
\usepackage{lipsum,multicol}
\usepackage{xcolor}
\usepackage{tikz}
\usepackage{listings}
\usepackage{enumitem}
\usepackage{hyperref}
\usepackage{amsfonts}
\usepackage{wrapfig}
\usepackage{subcaption} 
\usepackage{adjustbox}
\usepackage{colortbl}
\usepackage{fancybox}
\usepackage{multirow}
\usepackage[normalem]{ulem}
\useunder{\uline}{\ul}{}
\usepackage{enumitem}

\usepackage{booktabs}

\setcitestyle{maxnames=2, minnames=1}

\newcommand{\ie}{\textit{i.e.,}\xspace}

\newtcolorbox{boxK}{
    fontupper = \small,
    sharpish corners, %
    boxrule = 0pt,
    toprule = 0pt, %
}

\newcommand*\circled[1]{\tikz[baseline=(char.base)]{
            \node[shape=circle,draw,inner sep=0.5pt] (char) {#1};}}

\newcommand{\llms}{LLMs\xspace}
\newcommand{\llm}{LLM\xspace}

\newcommand{\llmc}{LLMc\xspace}

\begin{document}
\title{ 
Towards Comprehensive Benchmarking Infrastructure for LLMs In Software Engineering
}

\settopmatter{authorsperrow=4}

\author{Daniel Rodriguez-Cardenas}
\email{dhrodriguezcar@wm.edu}
\affiliation{%
  \institution{William \& Mary}
  \city{Williamsburg}
  \state{Virginia}
  \country{USA}
}

\author{Xiaochang Li}
\email{xli59@wm.edu}
\affiliation{%
  \institution{William \& Mary}
  \city{Williamsburg}
  \state{Virginia}
  \country{USA}
}

\author{Marcos Macedo}
\email{marcos.macedo@queensu.ca}
\affiliation{%
  \institution{Queen's University}
  \city{Kingston}
  \state{Ontario}
  \country{Canada}
}

\author{Antonio Mastropaolo}
\email{amastropaolo@wm.edu}
\affiliation{%
  \institution{William \& Mary}
  \city{Williamsburg}
  \state{Virginia}
  \country{USA}
}

\author{Dipin Khati}
\email{dkhati@wm.edu}
\affiliation{%
  \institution{William \& Mary}
  \city{Williamsburg}
  \state{Virginia}
  \country{USA}
}
\author{Yuan Tian}
\email{y.tian@queensu.ca}
\affiliation{%
  \institution{Queen's University}
  \city{Kingston}
  \state{Ontario}
  \country{Canada}
}

\author{Huajie Shao}
\email{hshao@wm.edu}
\affiliation{%
  \institution{William \& Mary}
  \city{Williamsburg}
  \state{Virginia}
  \country{USA}
}

\author{Denys Poshyvanyk}
\email{dposhyvanyk@wm.edu}
\affiliation{%
  \institution{William \& Mary}
  \city{Williamsburg}
  \state{Virginia}
  \country{USA}
}

\renewcommand{\shortauthors}{Rodriguez-Cardenas et al.}

\begin{abstract}

Large language models for code are advancing fast, yet our ability to evaluate them lags behind. Current benchmarks focus on narrow tasks and single metrics, which hide critical gaps in robustness, interpretability, fairness, efficiency, and real-world usability. They also suffer from inconsistent data engineering practices, limited software engineering context, and widespread contamination issues. To understand these problems and chart a path forward, we combined an in-depth survey of existing benchmarks with insights gathered from a dedicated community workshop. We identified three core barriers to reliable evaluation: the absence of software-engineering-rich datasets, overreliance on ML-centric metrics, and the lack of standardized, reproducible data pipelines. Building on these findings, we introduce BEHELM, a holistic benchmarking infrastructure that unifies software-scenario specification with multi-metric evaluation. BEHELM provides a structured way to assess models across tasks, languages, input and output granularities, and key quality dimensions. Our goal is to reduce the overhead currently required to construct benchmarks while enabling a fair, realistic, and future-proof assessment of LLMs in software engineering.

\end{abstract}



\maketitle

\section{Introduction}\label{sec:introduction}

Large Language Models for code (LLMc) are reshaping software engineering workflows. Their growing capabilities allow developers to rely on them for tasks such as code completion~\cite{austin2021program, hendrycks2021, chen2021evaluating,White.MSR.2015,Ciniselli.MSR}, summarization~\cite{Hussain2020DeepTL, leclair_ensemble_2021,Moran.SANER.2022}, code review assistance \cite{Mastropaolo.TSE, Mastropaolo.ICSE, Tufano.ICSE.2021, Tufano.ICSE.2022}, traceability \cite{Moran.Traceability.ICSE}, translation \cite{Nguyen:ICSE15}, test generation~\cite{white_reassert_2020, Raychev2014CodeCW,Watson.Asserts.ICSE}, code clode detection~\cite{White.ASE2016}, and bug fixing~\cite{Tufano2019LearningBugFixes,zhou_devign_nodate,Tufano.ICSE19.Changes,ASE.2018,Zimin.Sequencer,CanWeFix,8668043}. Tools like GitHub Copilot~\cite{wermelinger2023using} and Google's CodeBot~\cite{ rosenberg2020codebots} show how these models can support everyday development at scale, from writing small snippets to generating full functions. As a result, LLMc now plays a concrete role in improving developer efficiency and accelerating software production.

This rapid adoption has increased the need for a reliable and standardized evaluation of model performance \cite{liu2023code,liu2023improving,xu_systematic_2022,chen2021evaluating}. Current assessment methods are heavily focused on accuracy metrics \cite{lu_codexglue_2021,cassano2022multiplescalableextensibleapproach,openai_codex} and robustness metrics \cite{wang_recode_2022}. However, these metrics reveal little about important factors such as \emph{interpretability, efficiency, fairness, bias, and resilience to real-world complexity}. In practice, we still lack a clear understanding of how \llmc behaves in various software engineering scenarios~\cite{liang_holistic_2022}. Existing benchmarks do not capture the full range of properties that influence code quality, leaving critical blind spots in model evaluation.

These limitations stem from several persistent challenges in building and maintaining evaluation resources for \llmc. Benchmark creation often requires extensive data collection and cleaning, careful labeling, and ongoing maintenance to keep up with model progress. Many datasets focus on simplified or isolated tasks, and few offer the kind of rich context on which developers rely in real projects. The problems of data leakage, inconsistent preprocessing, and unclear provenance further complicate a meaningful comparison between studies.

To address these challenges, this article synthesizes community perspectives from a workshop developed to identify the main gaps in \llmc evaluation and proposes a novel framework to address them. We highlight shortcomings in current datasets, metrics, and data engineering practices and provide recommendations for developing a more robust, comprehensive benchmarking infrastructure tailored to fundamental software engineering needs.

This paper contributes: (1) synthesis of workshop discussions identifying critical infrastructure gaps\cite{benchmarking-repo-70FA}; (2) comprehensive survey of current benchmarks in SE; (3) proposes a novel holistic benchmark for LLMs in SE; (4) concrete recommendations for community infrastructure development.

\section{Current Benchmark Landscape}

From our survey, we focus on five main SE tasks (\ie  (1) code generation, (2) bug-fixing, (3) vulnerability detection, (4) test generation, (5) Agent and Multi-Task) and the most relevant benchmarks reported at the leading conferences (\ie ICSE, ASE, FSE, TOSEM).

\textbf{(1) Code Generation Benchmarks.} \emph{HumanEval} (164 problems) and \emph{MBPP} (974 problems) remain widely used despite saturation: the top models now exceed 90\% accuracy in HumanEval, making it ineffective for differentiating capabilities~\cite{chen2021evaluating}. \emph{APPS} offers harder competitive-programming tasks, but still focuses on algorithmic correctness rather than software engineering~\cite{hendrycks2021}.

\emph{LiveCodeBench} mitigates contamination by continuously collecting problems from LeetCode, AtCoder, and CodeForces (May 2023–present)~\cite{jain2024live}. Its 511+ tasks evaluate generation, self-repair, execution, and test prediction. Crucially, performance drops 15–20\% in problems released after model training cutoffs, revealing substantial contamination effects.

More realistic evaluations emerge in \textit{class-level} and \textit{repository-level} benchmarks. \emph{ClassEval} (100 tasks) tests the ability to generate full classes~\cite{du2023class}, while \emph{EvoCodeBench} (275 samples from 25 repositories) reduces data leakage from 41.47\% to 2.18\% and reports only 20.74\% Pass@1 for the best models~\cite{li2024evo}, underscoring the difficulty of real-world tasks.

Multi-lingual benchmarks expand the evaluation beyond Python: \emph{HumanEval-X} and \emph{MultiPL-E} span 5–18 languages~\cite{zheng2023codegeex,cassano2022multiplescalableextensibleapproach}, while \emph{DS-1000} provides 1,000 data-science–oriented problems in Python~\cite{Lai2022DS1000}.

\textbf{Workshop insight}: Small benchmarks ($\approx$200 examples) work well as “smoke tests’’ for identifying major failures—much like polls detect landslides—but are insufficient for fine-tuning or comprehensive capability assessment.

\textbf{(2) Bug Fixing and Program Repair.} \emph{SWE-bench} marks a major step forward with 2,294 real GitHub issues and ground-truth fixes from 12 Python repositories~\cite{jimenez2024swebenchlanguagemodelsresolve}. Concerns about infeasible cases prompted curated subsets such as \emph{SWE-bench Verified} (500 human-validated instances) and \emph{SWE-bench Lite} (300 curated tasks)~\cite{openai2024swe}. Yet top models that achieve over 70\% on SWE-bench Verified drop to just 23\% on SWE-bench Pro, which uses private commercial codebases and recent issues~\cite{scale2025swe}. This 47-point decline strongly indicates memorization rather than genuine repair capability.

Several extensions attempt to broaden or strengthen evaluation. \emph{SWE-bench Java} adds 500+ Java issues~\cite{zan2024swe}, \emph{SWE-bench Multimodal} incorporates 617 JavaScript tasks with visual components~\cite{sweetbench2024}, and \emph{OSS-bench} generates tasks dynamically from recent pull requests to mitigate leakage~\cite{jiang2025ossbenchbenchmarkgeneratorcoding}. Beyond these, \emph{SWT-bench} introduces dual evaluation—reproducing issues via generated tests and validating fixes with test-based filtering—which can double the precision of repair agents~\cite{mundler2025swtbenchtestingvalidatingrealworld}.

A key insight from the workshop is that data leakage persists even in “live’’ benchmarks because code is frequently copied between projects. World of Code analysis show that roughly 65\% of Python packages provide a repository, meaning that partial or duplicated code can reappear in multiple contexts~\cite{ma2021world}. Temporal filtering alone cannot prevent leakage; comprehensive supply-chain analysis is required in software.

\textbf{(3) Vulnerability Detection and Repair.} \emph{CVEfixes} compiles 5,867 historical C/C++ CVEs (1999–2024) for statement-level vulnerability localization but lacks executable exploits to confirm real-world exploitability~\cite{ahmed2025}. \emph{SecBench.JS} addresses this gap with 600 JavaScript vulnerabilities paired with working proof-of-concept exploits and test oracles~\cite{secbench2025}. Its two-sided design—providing both vulnerable and fixed versions—revealed 18 incorrect patches during validation, underscoring its utility for assessing patch quality. \emph{SEC-bench} further automates benchmark creation using multi-agent pipelines~\cite{secbench2025}. From 898 seed CVEs, it generated 200 fully verified instances with reproducible exploits and validated fixes at only $\$0.87$ per instance. Even the best LLMs remain limited, achieving just 49\% success in PoC generation and 34\% in patching.

\textbf{(4) Test Generation Benchmarks.} \emph{TestGenEval} provides 68,647 tests spanning 1,210 code–test pairs from 11 Python repositories, enabling evaluation of initial test authoring, suite completion, and coverage improvement in real codebases~\cite{jain2025testgenevalrealworldunit}. \emph{ULT} offers 3,909 high-complexity Python functions with strong contamination controls~\cite{9969493}. The high correlation between test-generation and code-generation performance ($r=0.79$, $p=0.002$) suggests that test generation may serve as a reliable proxy for general coding ability. Complementing these, \emph{SWT-bench} shows that code-repair agents can outperform dedicated test-generation tools~\cite{mundler2025swtbenchtestingvalidatingrealworld}. With an LLM-optimized diff format, Pass@1 improves from 3.3\% to 9.4\% (and to 20.3\% at Pass@5), although reliably reproducing failure tests remains a major challenge.

\textbf{(5) Agent and Multi-Task Benchmarks.} \emph{CodeXGLUE} offers 14 datasets covering 10 tasks in code-to-code, text-to-code, code-to-text, and text-to-text settings~\cite{lu_codexglue_2021}, but its scenarios and metrics remain limited for modern SE needs. More specialized efforts include \emph{CrossCodeBench}, which evaluates cross-task generalization~\cite{niu2023crosscodebenchbenchmarkingcrosstaskgeneralization}; \emph{ReCode}, which targets robustness~\cite{wang_recode_2022}; and \emph{Galeras}, which studies interpretability through causal analysis~\cite{10336302}. Holistic evaluations, such as Yuan et al.’s study of 10 instruction-tuned LLMs across code-understanding and generation tasks~\cite{yuan2023}, still assess only narrow slices of model behavior using basic metrics.

Workshop discussions emphasized that existing benchmarks fail to capture real usage of modern AI-assisted development tools (e.g., Cursor, Copilot, Gemini CLI). Evaluating agentic systems requires process-aware milestone-based assessments that track progress through development stages rather than relying solely on binary success or failure.

\section{Critical Gaps in Benchmarking Infrastructure}
This section emphasizes the identified gaps according to our paper survey and the viewpoint of the workshop participants'.

\textbf{Gap 1: Lack of Specialized SE Datasets.} Most \llm datasets focus on source code, input/output examples, and natural language, while neglecting essential software engineering attributes. For program synthesis evaluation, researchers construct datasets from model input/output cases but ignore SE-specific attributes that are crucial for realistic evaluation. Below, we list the missing SE attributes: 

\begin{itemize}
  \item \textbf{Software artifacts}: Repository structure, build configurations, dependency manifests
  \item \textbf{Process metadata}: Commit histories, code review comments, issue discussions, pull request interactions
  \item \textbf{Quality metrics}: Code complexity, maintainability indices, technical debt indicators
  \item \textbf{Context information}: Project documentation, API specifications, architectural constraints
\end{itemize}

Natural language and source code dominate evaluation datasets, while software artifacts and repository metadata---containing rich contextual information essential for realistic code generation---are seldom considered. This narrow focus produces benchmarks that evaluate syntactic correctness without assessing whether generated code integrates properly into real software projects.

Workshop participants emphasized that benchmarks should incorporate the full software development lifecycle (SDLC), not just isolated code generation. Real developers work with existing codebases, navigate complex dependencies, and follow project-specific conventions---none of which current benchmarks adequately capture~\cite{ma2021world}.

\textbf{Gap 2: ML Metrics vs. SE-Specific Evaluation} Current evaluation relies heavily on machine learning metrics (accuracy, precision, recall, F1) and text similarity measures (BLEU, CodeBLEU) that fail to capture software engineering quality.

\begin{table}[h]
\centering
\resizebox{\linewidth}{!}{%
\begin{tabular}{lll}
\textbf{Metric Category} & \textbf{Common Metrics} & \textbf{SE Limitations} \\
\midrule
\textbf{Classification} & Accuracy, Precision, Recall, F1 & Binary success/failure ignores partial \\
& & correctness and intermediate progress \\
\textbf{Text Similarity} & BLEU,  CodeBLEU & Word overlap doesn't capture semantic \\
& & equivalence or multiple valid solutions \\
\textbf{Test-Based} & AvgPassRatio, Test Pass Rate & Passing tests doesn't guarantee code \\
& & quality, security, or maintainability \\
\textbf{Embedding-Based} & BERTScore, CodeBERTScore & Similarity in embedding space doesn't \\
& & ensure functional correctness \\
\textbf{Correlation} & Kendall-Tau, Spearman, Pearson & Statistical correlation doesn't capture \\
& & causal relationships or interpretability \\
\bottomrule
\end{tabular}
}
\caption{Limitations of current evaluation metrics for code generation}
\label{tab:metrics}
\vspace{-2em
}
\end{table}
Test passing alone is insufficient for evaluating model performance. Many problems admit multiple valid solutions, yet most benchmarks provide only a single reference implementation. The “SWE-bench Illusion’’ study shows that models can reach 76\% context-free accuracy with high $n$-gram overlap, suggesting they rely on memorization rather than genuine reasoning~\cite{swebench_illusion}. Moreover, performance rankings closely follow benchmark curation—SWE Verified (34.9\%) > SWE Full (28.7\%) > SWE Extra (18.2\%)—which further indicates that models may be learning curated solution patterns instead of developing broad problem-solving ability~\cite{swebench_illusion}.

We identify several evaluation dimensions that are largely missing from current SE benchmarks, including \textbf{interpretability} to understand and justify a model’s code-generation process, \textbf{efficiency} for resource usage, runtime behavior, and algorithmic complexity, \textbf{bias and fairness} whether the model produces biased or inequitable code, \textbf{robustness} to measure the performance under distribution shift, adversarial inputs, or rare edge cases, \textbf{maintainability} for readability, modularity, and adherence to best practices, and \textbf{security} including vulnerability patterns, unsafe API usage, and injection risks.

Workshop participants emphasized the need for intent-based evaluation, which acknowledges semantically equivalent solutions that differ syntactically~\cite{ma2021world}. When benchmarks rely on a single reference solution—as most currently do—models are incentivized to overfit to specific forms and patterns rather than demonstrating genuine generalization.

\textbf{Gap 3: Lack of Standardized Data Engineering Pipelines} Existing benchmarks are constructed in highly heterogeneous ways, making it difficult to compare results fairly and creating substantial overhead. As Watson et al.\ argue, standardized workflows should explicitly document the data source, data type, and all preprocessing steps for both training and testing sets~\cite{watson2022}. In practice, however, each research group builds its own benchmark from scratch, leading to duplicated infrastructure work, uneven quality standards, limited reproducibility due to undocumented preprocessing, and significant resource costs—often 50–75\% of a paper’s effort is spent solely on dataset construction\cite{ma2021world}.

Workshop participants repeatedly highlighted that this burden of benchmark construction is becoming unsustainable. The SecBench automation study found that 75\% of project time was consumed by dataset creation~\cite{secbench2025}. Such inefficiency suppresses diversity and innovation: researchers are forced to choose between developing new evaluation methods or reusing existing but often inadequate benchmarks. Participants emphasized the need for data provenance tracking (to know where each datapoint originates and whether it appeared in training), rigorous quality-control workflows (including inter-rater reliability checks, expert review, and contamination detection), and automated pipelines for extraction, cleaning, deduplication, and validation. They also stressed the importance of standardized data formats—covering schemas, evaluation scripts, and result-reporting conventions—as well as robust versioning mechanisms that allow benchmarks to evolve without compromising reproducibility.

\section{Benchmarking Infrastructure Development}
In this section, we first discuss our findings from the workshop and survey results, then we formulate an infrastructure for a holistic evaluation framework for \llms in SE.
\begin{figure*}[!h]
	\centering
	\includegraphics[width=0.9\textwidth]{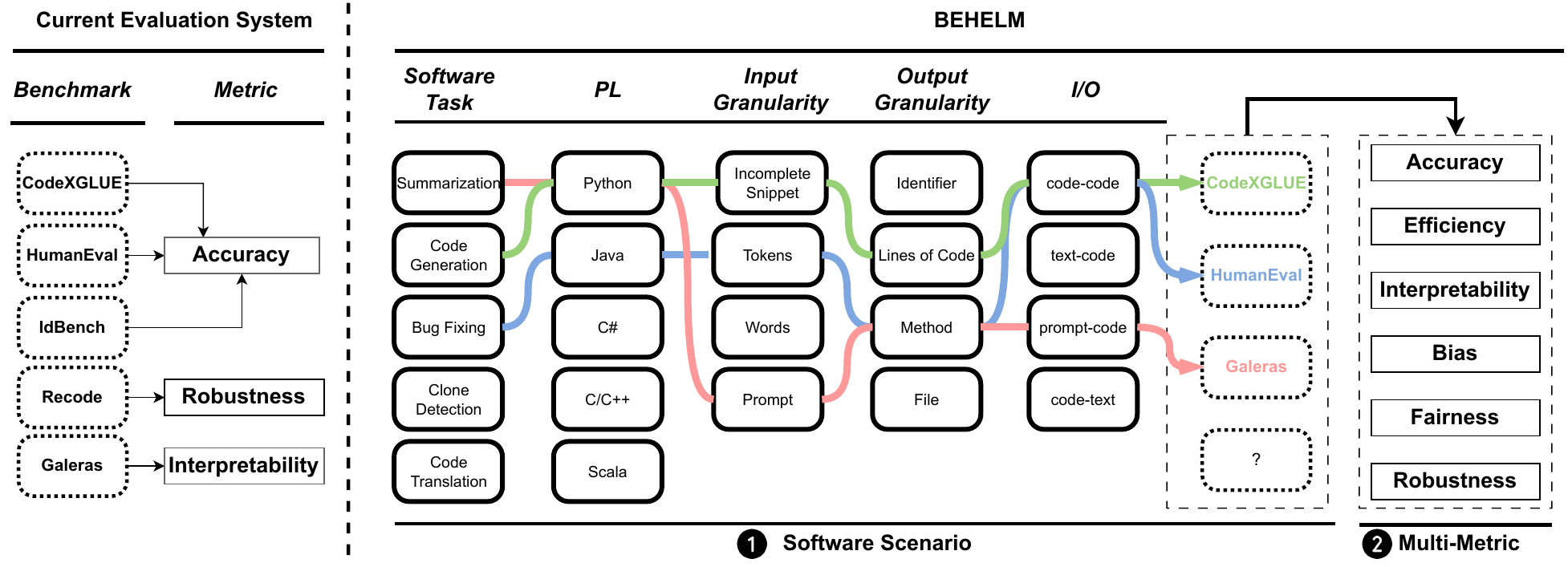}
    \vspace{-0.2in}
	\caption{Current Evaluation System vs. Infrastructure for Holistic Evaluation of \llmc .}
        \label{fig:behelm}
\end{figure*}

The workshop discussions revealed a clear need for a holistic next-generation benchmark ecosystem for SE-focused LLMs. The participants agreed that the current benchmarks are fragmented, narrow in scope, and insufficient to evaluate modern AI-driven development workflows. Four gaps emerged as the central motivation.

First, the \textbf{community lacks standardized data pipelines}: researchers need a \emph{unified platform} that provides consistent dataset formats, robust provenance tracking, contamination detection, and high-quality annotation workflows. Second, \textbf{evaluation must expand beyond accuracy} to a multi-metric framework that captures interpretability, robustness, fairness, efficiency, security, and partial progress—while supporting lifecycle and milestone-based assessment for agentic systems. Third, \textbf{sustainable benchmarking} requires an automated infrastructure, including provenance resources, semantic deduplication, and RL-driven mechanisms to generate diverse solutions and prevent overfitting. Finally, participants stressed the importance of \textbf{community coordination}, shared standards, and collaborative development to avoid redundant efforts and ensure that benchmarks remain relevant, diverse, and grounded in real-world SE practice.

Together, these points motivate the construction of a comprehensive community-driven benchmark capable of evaluating LLMs across the full spectrum of software engineering tasks, metrics, and development stages.

\subsection*{BEHELM Infrastructure for Holistic Evaluation}

The BEHELM (Benchmarking for Holistic Evaluation of Large Language Models for Code) framework represents a comprehensive infrastructure designed to evaluate Large Language Models for code (LLMc) beyond traditional accuracy metrics. The framework, illustrated in Figure \ref{fig:behelm}, comprises two main components: the current evaluation system and the proposed BEHELM infrastructure.

The left side of the figure depicts the capabilities (and limitations) of existing benchmarking approaches. Traditional benchmarks such as CodeXGLUE, HumanEval, IdBench, Recode, and Galeras are shown mapping to narrow evaluation metrics. Specifically, CodeXGLUE and HumanEval primarily focus on \textbf{accuracy} metrics, IdBench is limited to accuracy evaluation, Recode emphasizes \textbf{robustness} assessment, and Galeras addresses \textbf{interpretability}. This approach demonstrates that current evaluation systems typically measure a single metric, failing to account for how models perform across multiple dimensions. Each benchmark operates independently with its own canonical metric, lacking a unified framework that enables comprehensive, multi-faceted evaluation of LLMc capabilities across diverse software engineering scenarios.

The right side of the Figure~\ref{fig:behelm} presents BEHELM's innovative approach through two interconnected components: \textbf{Software Scenario} specification (labeled as \circled{1}) and \textbf{Multi-Metric} evaluation (labeled as \circled{2}).

The \textbf{Software Scenario} component introduces a systematic taxonomy for defining evaluation contexts. Each scenario is decomposed into five fundamental properties that control and standardize the evaluation environment: \textbf{Software Task}, which includes activities such as summarization, code generation, bug fixing, clone detection, and code translation; \textbf{Programming Language (PL)}, supporting multiple languages, including Python, Java, C\#, C/C++, and Scala; \textbf{Input Granularity}, which categorizes the level of detail in the input representation (incomplete snippet, tokens, words, or prompt); \textbf{Output Granularity}, defining the expected level of output detail (identifier, lines of code, method, or file); and \textbf{I/O type}, specifying the interaction modality between input and output (code-to-code, text-to-code, code-to-text, or prompt-to-code). This structured decomposition enables researchers to explicitly state what software properties require evaluation based on real-world use cases, ensuring that each LLMc is assessed in identical controlled settings for fair comparison.

The \textbf{Multi-Metric} evaluation component represents BEHELM's holistic assessment approach, measuring six critical dimensions: \textbf{accuracy}, \textbf{efficiency}, \textbf{interpretability}, \textbf{bias}, \textbf{fairness}, and \textbf{robustness}. Unlike traditional benchmarks that evaluate a single metric in isolation, BEHELM creates a comprehensive evaluation space defined by the Cartesian product of software scenarios and metrics. This multidimensional framework enables practitioners to understand not only whether a model generates correct code, but also whether it does so efficiently, transparently, equitably, and reliably across diverse conditions. The figure illustrates how specific benchmarks (CodeXGLUE, HumanEval, Galeras) can be positioned within this framework, with some scenarios lacking comprehensive metric coverage (indicated by question marks), highlighting gaps that BEHELM addresses through its unified infrastructure.

\section{Conclusions and Future Work}

The workshop identified three \textit{critical gaps} that hinder comprehensive LLM evaluation for software engineering: first, the lack of specialized SE datasets that incorporate software artifacts and process metadata; second, inadequate metrics that overemphasize accuracy while neglecting interpretability, efficiency, fairness, and robustness; and third, the absence of standardized data engineering pipelines for benchmark creation.

We formulated a holistic framework to address \textbf{three critical research questions} : (1) How to create realistic code editing/refactoring benchmarks? (2) How to perform intent-based evaluation beyond test-passing? (3) How to benchmark modern AI-assisted development tools in real usage scenarios?

We plan for BEHELM to bring current benchmarks together while enabling the creation of new ones, offering a coordinated framework for evaluating LLMs in software engineering. As a central hub, BEHELM will provide standardized tools for profiling, comparing, and debugging model behavior across SE tasks.

\section{Acknowledgments}

We want to thank the workshop participants for their active engagement, insightful discussions, and constructive feedback. Their perspectives greatly enriched the ideas presented in this work and helped shape several of the directions we explore.

\bibliographystyle{ACM-Reference-Format-num}
\bibliography{utils/references}

\end{document}